\documentclass[aps,prb,amsmath,amssymb,reprint,superscriptaddress]{revtex4-2}

\usepackage{graphicx}
\usepackage{amsmath,amssymb}
\usepackage{xcolor}
\usepackage{color,soul}

\usepackage{geometry}
\usepackage{hyperref}
\usepackage{listings}
\usepackage{xcolor}
\usepackage{url}
\usepackage{caption}
\usepackage{float}
\usepackage{listings}
\usepackage{xcolor}

\begin{document}

\title{Revisiting and Accelerating the Basin Hopping Algorithm for Lennard–Jones Clusters: Adaptive and Parallel Implementation in Python}

\author{Oliver Carmona}
\affiliation{Centro de Investigaciones en \'Optica A.C., Loma del Bosque 115, Lomas del Campestre, Leon, 37150, Guanajuato, Mexico} 
\author{P. L. Rodr\'iguez-Kessler}
\email{plkessler@cio.mx}
\affiliation{Centro de Investigaciones en \'Optica A.C., Loma del Bosque 115, Lomas del Campestre, Leon, 37150, Guanajuato, Mexico} 
\author{S. Salazar-Colores}
\affiliation{Centro de Investigaciones en \'Optica A.C., Loma del Bosque 115, Lomas del Campestre, Leon, 37150, Guanajuato, Mexico} 
\author{Alvaro Mu\~{n}oz-Castro}
\affiliation{Facultad de Ingenier\'ia, Universidad San Sebasti\'an, Bellavista 7, Santiago, 8420524, Chile.}

\date{\today}

\begin{abstract}
We present an adaptive and parallel implementation of the Basin Hopping (BH) algorithm for the global optimization of atomic clusters interacting via the Lennard–Jones (LJ) potential. The method integrates local energy minimization with adaptive step-size Monte Carlo moves and simultaneous evaluation of multiple trial structures, enabling efficient exploration of complex potential energy landscapes while maintaining a balance between exploration and refinement. Parallel evaluation of candidate structures significantly reduces wall-clock time, achieving nearly linear speedup for up to eight concurrent local minimizations. This framework provides a practical and scalable strategy to accelerate Basin Hopping searches, directly extendable to ab initio calculations such as density functional theory (DFT) on high-performance computing architectures.
\end{abstract}


\maketitle

\section{Introduction}

The structural optimization of atomic clusters remains a central challenge in computational physics and chemistry due to their highly complex potential energy surfaces (PES) with numerous local minima. This complexity is especially pronounced when interatomic interactions are modeled with non-trivial potentials, such as the Lennard-Jones (LJ) form.

Although numerous structure search methods for atomic clusters have been developed, most share the common goal of efficiently exploring the PES. Given the extensive literature, we highlight a representative selection of methods \cite{doi:10.1021/ct050093g,https://doi.org/10.1002/jcc.10407,Tong_2019,doi:10.1021/acs.jctc.6b00556,doi:10.1021/acs.jctc.8b00772,C5CP04060D,https://doi.org/10.1002/anie.202500393,https://doi.org/10.1002/qua.26553}, while acknowledging that many important contributions are discussed in the following reviews \cite{https://doi.org/10.1002/qua.26553,D0CP06179D}.

In this brief report, we focus on the Basin Hopping (BH) algorithm, providing a concise overview of its foundational principles and highlighting strategies to enhance its efficiency. BH is one of the most effective methods for navigating rugged PES landscapes. By alternating between Monte Carlo perturbations and local minimization, BH maps the high-dimensional PES into basins of attraction, allowing the search to focus on energetically relevant regions rather than the full PES. Its efficiency, however, depends strongly on key parameters such as the perturbation step size and the robustness of the local optimization routine.

Basin Hopping has been extensively studied in seminal works by Wales and Doye,\cite{doi:10.1021/jp970984n} particularly in the context of Lennard-Jones clusters, which serve as prototypical systems for studying cluster energetics and structural motifs. Despite this, no publicly available, standardized Python implementation exists that is both user-friendly and adaptable for modern research and teaching purposes.

Among Lennard-Jones clusters with fewer than 100 atoms, certain configurations, such as LJ$_{38}$, present particularly challenging energy landscapes. These so-called "hard clusters" are widely used as benchmarks for structural search algorithms, reinforcing the role of Lennard-Jones systems as reference models in global optimization studies.\cite{doi:10.1021/acs.jctc.4c00365}

Such complex landscapes underscore the need for robust and flexible optimization frameworks that can efficiently explore low-energy configurations in a computationally feasible manner. Addressing this need, we present a Python implementation of BH designed to be accessible, modifiable, and computationally efficient. Our version extends the classical BH framework through two key improvements:
\begin{itemize}
\item \textbf{Adaptive Step Size:} The amplitude of random perturbations is dynamically adjusted based on the recent acceptance rate, enhancing sampling efficiency across diverse regions of the PES.
\item \textbf{Parallel Local Minimization:} Multiple perturbed candidates are minimized concurrently at each BH step, and the structure with the lowest energy is selected for the Metropolis criterion. This improves convergence and reduces the probability of prolonged trapping in suboptimal minima.
\end{itemize}

Although classical Basin Hopping has been applied in DFT calculations,\cite{doi:10.1021/acs.jpcc.6b11968,PhysRevB.84.193402,doi:10.1021/acs.jpca.5b04015,doi:10.1021/ci400224z} such approaches remain computationally demanding. Our code can also serve as a foundation for integrating machine-learned potentials, such as neural networks trained on quantum mechanical data, enabling more efficient structural exploration with near-DFT accuracy.\cite{doi:10.1021/acs.jctc.4c01257,10.1063/1674-0068/cjcp2309083,doi:10.1021/acs.jctc.6b00994}

\section{Methodology}

\subsection{Lennard-Jones Potential}

The Lennard-Jones (LJ) potential is used to model the pairwise interaction between atoms. The total energy of an $N$-atom cluster is computed as:

\begin{equation}
E_{\text{LJ}} = 4\epsilon \sum_{i<j} \left[ \left( \frac{\sigma}{r_{ij}} \right)^{12} - \left( \frac{\sigma}{r_{ij}} \right)^6 \right],
\end{equation}

where \( r_{ij} \) is the distance between atoms \( i \) and \( j \), and \( \epsilon = \sigma = 1.0 \) in reduced LJ units.

\subsection{Initial Structures and Input}

The algorithm reads the first geometry from a \texttt{.molden} file. The file must contain Cartesian coordinates for all atoms in the structure. These coordinates are flattened into a 1D array suitable for optimization.

\subsection{Algorithm Overview}

The BH algorithm proceeds as follows:

\begin{enumerate}
    \item \textbf{Perturbation:} A random displacement is applied to all atoms within a cube of side $2 \times \texttt{stepsize}$.
    \item \textbf{Parallel Local Optimization:} $n$ candidates are minimized in parallel using the L-BFGS-B algorithm.
    \item \textbf{Selection:} The best candidate is compared to the current structure.
    \item \textbf{Metropolis Criterion:} If the candidate has lower energy or passes a probabilistic test at $T = 1.0$, it is accepted.
    \item \textbf{Adaptation:} Every 10 steps, the acceptance rate is used to adjust the \texttt{stepsize} to approach a target rate of 50\%.
\end{enumerate}

\subsection{Parallel Evaluation of Candidates}

The local minimizations of trial candidates are distributed using Python’s \texttt{multiprocessing.Pool}. This allows multiple CPU cores (or, in high-performance computing environments, multiple nodes) to evaluate independent structures simultaneously, substantially reducing wall-clock time while maintaining stochastic sampling.

\begin{figure}[H]
    \centering
    \includegraphics[width=0.47\textwidth]{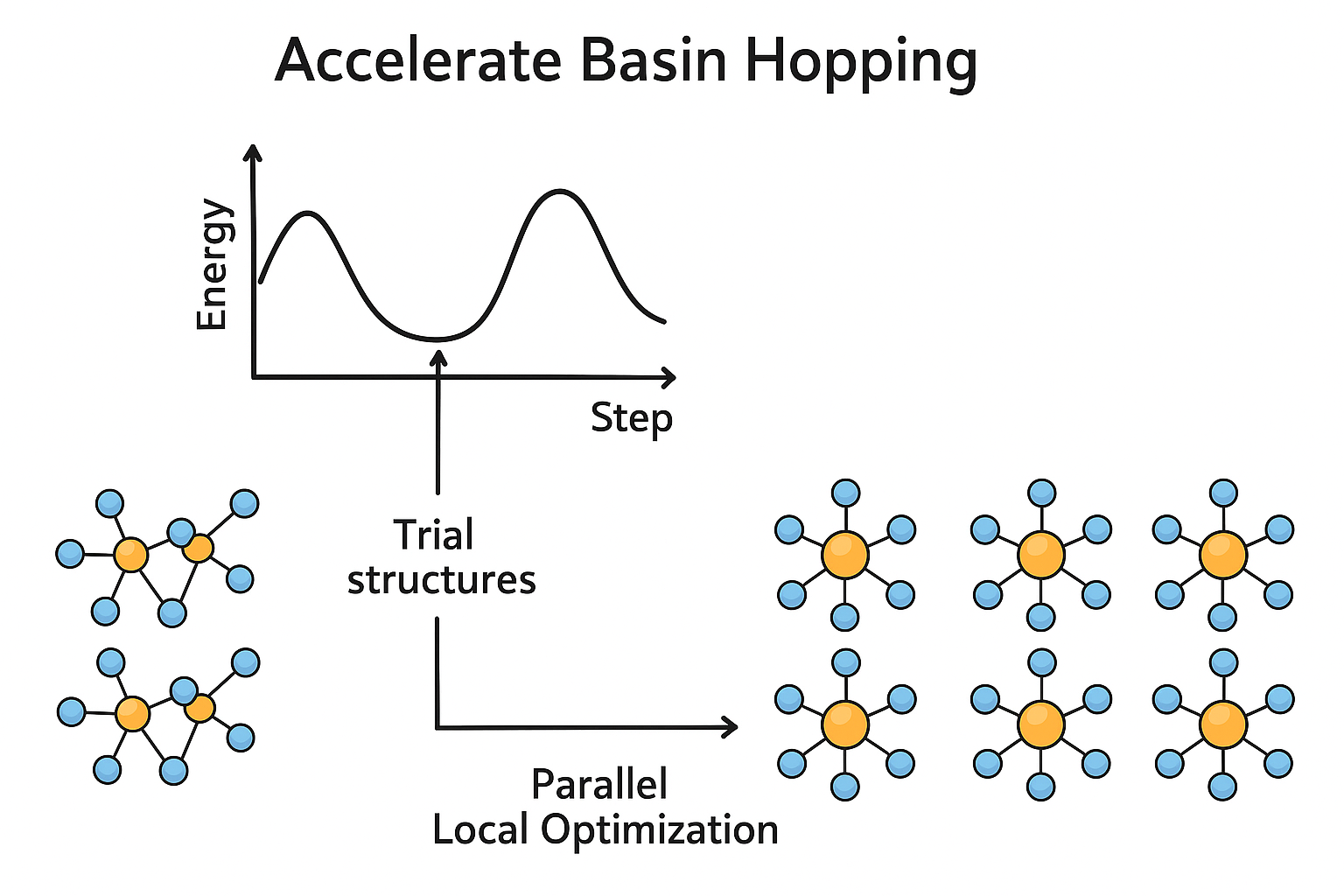}
    \caption{Schematic representation of the accelerated Basin Hopping workflow. Multiple trial structures are generated and locally optimized in parallel, enabling faster identification of low-energy configurations.}
    \label{fig:schematic}
\end{figure}

A schematic overview of the accelerated Basin Hopping workflow is shown in Fig.~\ref{fig:schematic}, illustrating how multiple trial structures are simultaneously minimized to identify low-energy configurations more efficiently.

\subsection{Implementation Notes}

The full code is written in Python and structured as a single script, \texttt{basin\_adapt.py}. Key dependencies are:

\begin{itemize}
    \item \texttt{numpy} – numerical operations,
    \item \texttt{scipy.optimize} – local minimizer (L-BFGS-B),
    \item \texttt{multiprocessing} – parallelism.
\end{itemize}

\subsection{Running the Code}

To execute the script, use the following command:

\begin{lstlisting}[language=bash]
python basin_adapt.py input.molden
\end{lstlisting}

The script will generate a file \texttt{input\_accepted.molden} with all accepted structures.

\section{Results}

The use of adaptive step sizes allows the Basin Hopping (BH) algorithm to dynamically balance exploration and evaluation. When the acceptance rate is too low, the step size decreases, improving sampling near the current configuration. Conversely, when the acceptance rate is too high, the step size increases, promoting broader exploration of the potential energy surface.

Simultaneous evaluation of multiple trial structures significantly reduces wall-clock time, as several candidate moves are processed concurrently. This is particularly beneficial for medium to large clusters, where each local optimization can be computationally expensive.

\begin{figure}[H]
\centering
\includegraphics[width=0.47\textwidth]{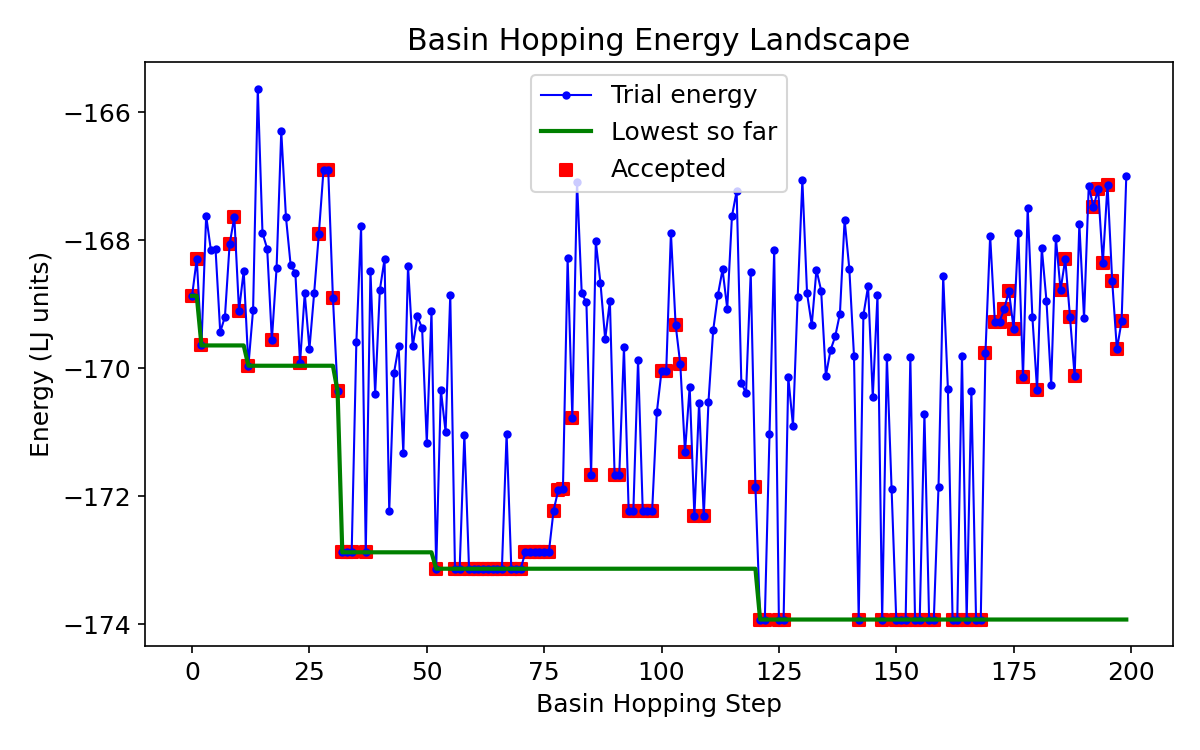}
\caption{Basin Hopping energy landscape for a Lennard-Jones cluster. Trial energies (blue), accepted configurations (red), and the running lowest energy (green) are shown as a function of the BH step.}
\label{fig:landscape}
\end{figure}

Figure~\ref{fig:landscape} presents the BH energy landscape over 200 steps. The blue line represents the trial energies at each step, while red squares indicate accepted configurations. The green curve tracks the lowest energy found so far. The plot highlights the stochastic nature of the search, with frequent fluctuations in trial energies, yet shows that the algorithm steadily improves the minimum energy. Acceptance of uphill moves allows the search to escape shallow local minima, while adaptive control ensures efficient convergence toward deeper basins.

The method consistently converges toward low-energy minima and avoids long-term trapping in unfavorable basins. Accepted configurations, recorded in the output Molden file, can be further analyzed for structural motifs or refined with higher-level quantum calculations.

\begin{figure}[H]
\centering
\includegraphics[width=0.47\textwidth]{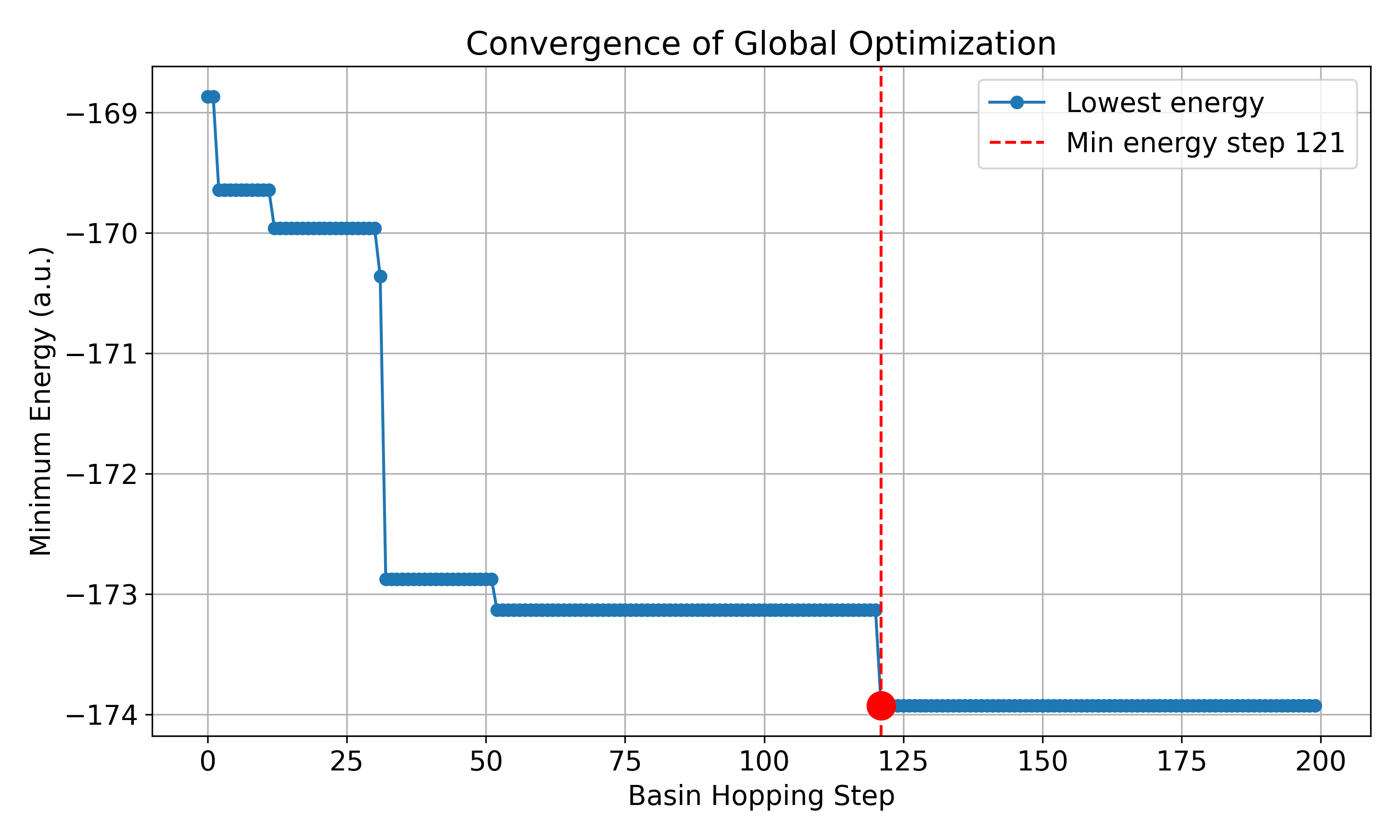}
\caption{Energy of the best accepted structure at each BH step. Adaptive control enables efficient descent.}
\label{fig:convergence}
\end{figure}

Figure~\ref{fig:convergence} shows the evolution of the lowest energy found during the BH run. Adaptive tuning of the step size prevents both stagnation and oversampling of a narrow region of configuration space.

\begin{figure}[H]
\centering
\includegraphics[width=0.47\textwidth]{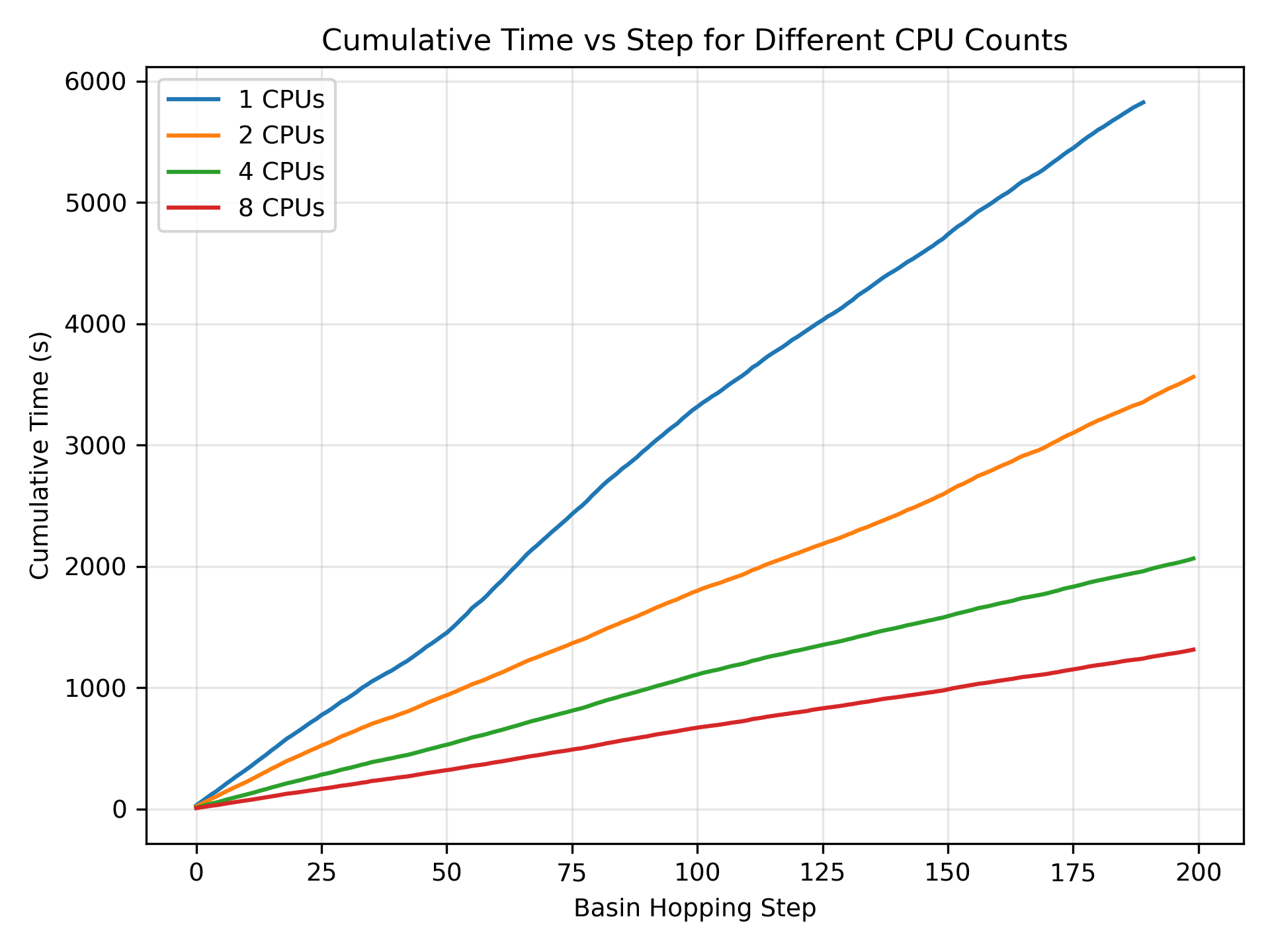}
\caption{Cumulative wall-clock time as a function of Basin Hopping step for different numbers of trial structures evaluated concurrently. Parallel evaluation substantially reduces runtime.}
\label{fig:cumulative_time}
\end{figure}

Figure~\ref{fig:cumulative_time} illustrates the cumulative wall-clock time as a function of BH steps for different numbers of trial structures evaluated concurrently. As expected, the total runtime decreases markedly with increased parallel evaluation. The improvement is particularly pronounced up to eight candidates per step, beyond which efficiency gains become modest due to synchronization overhead and the limited number of independent candidates per iteration. 

\begin{figure}[H]
\centering
\includegraphics[width=0.47\textwidth]{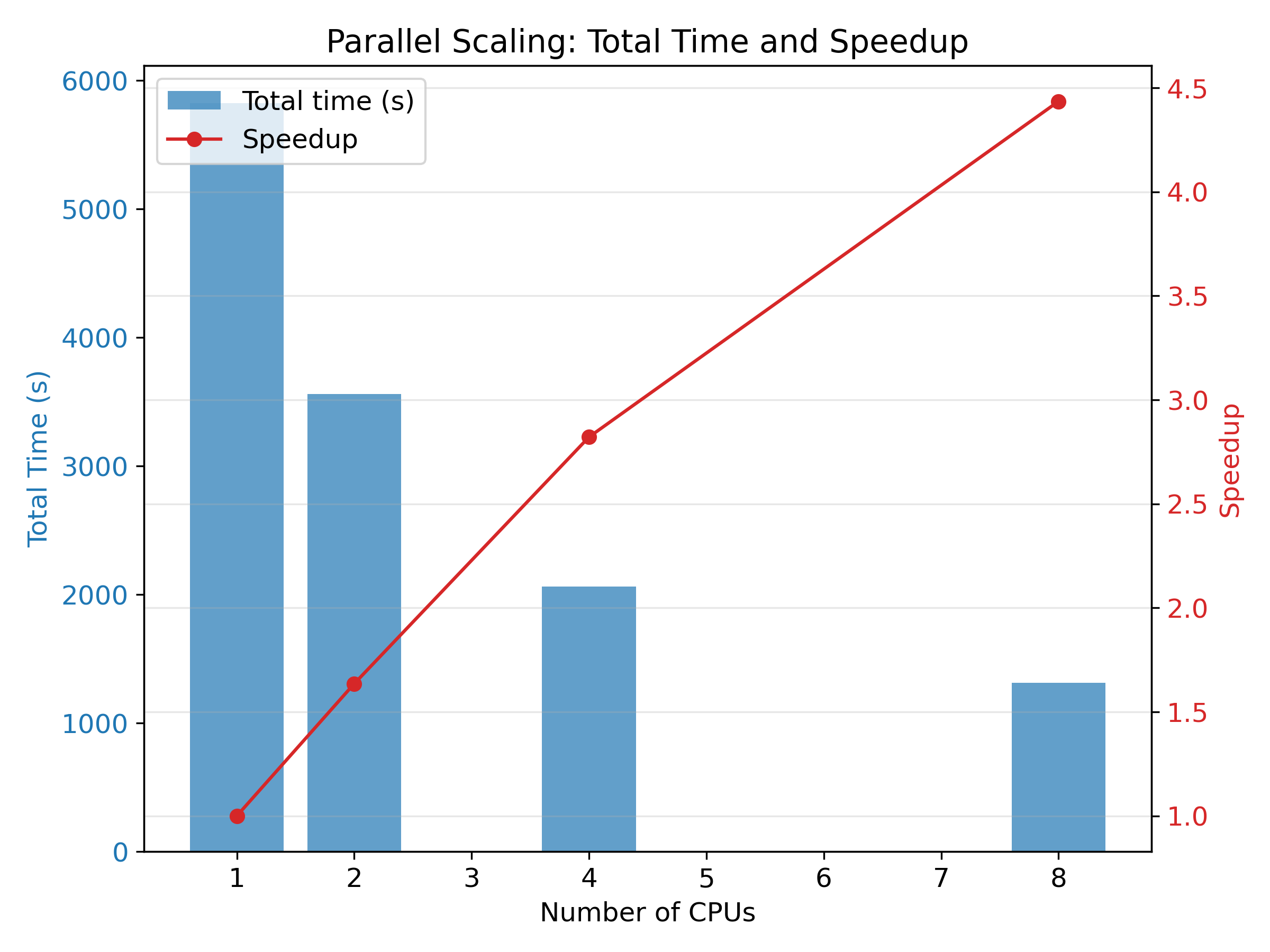}
\caption{Speedup and parallel efficiency of the Basin Hopping algorithm as a function of the number of trial structures evaluated concurrently. Efficiency approaches saturation beyond eight candidates per step due to communication and synchronization overhead.}
\label{fig:speedup_efficiency}
\end{figure}

Figure~\ref{fig:speedup_efficiency} summarizes the scaling performance relative to single-candidate evaluation, showing near-linear acceleration when up to eight structures are minimized concurrently. While further parallelization across more cores is possible in multithreading systems, these results effectively demonstrate a strategy for accelerating Basin Hopping steps through concurrent local minimizations. Such approaches can be directly extended to computationally demanding \textit{ab initio} calculations, where each candidate structure may be evaluated on a separate compute node in an HPC cluster rather than on individual cores.

\begin{figure}[H]
    \centering
    \includegraphics[width=0.47\textwidth]{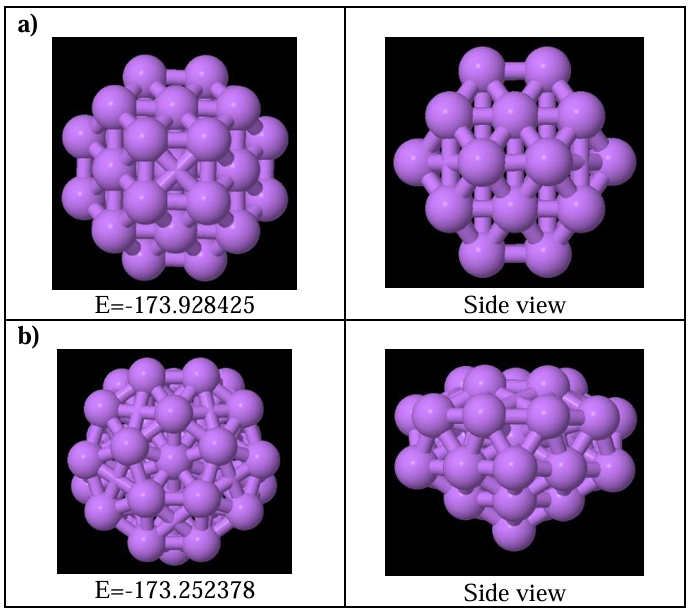}
    \caption{Representative low-energy configurations of the Lennard-Jones (LJ)\(_{38}\) cluster obtained from the Basin Hopping (BH) search. 
    (a) Global minimum: truncated octahedral (fcc-like) structure with near \(O_h\) symmetry. 
    (b) First low-lying isomer: incomplete icosahedral-derived motif with \(C_{5v}\) symmetry. 
    Both top and side views are shown for comparison.}
    \label{fig:structure}
\end{figure}

Finally, Fig.~\ref{fig:structure} displays representative low-energy configurations obtained from the Basin Hopping (BH) search of the Lennard-Jones (LJ)$_{38}$ cluster, a well-known benchmark system characterized by a particularly complex potential energy surface.\cite{D0CP06179D,doi:10.1021/acs.jctc.4c00365,10.1063/1.4901131} The global minimum corresponds to a compact, nearly spherical truncated octahedral structure with O$_h$ symmetry, while the closest low-lying isomer adopts a distorted icosahedral motif with C$_{5v}$ symmetry. The latter exhibits a shape reminiscent of a faceted diamond, formed by an incomplete icosahedral fragment. 

These optimized geometries exemplify the capability of the adaptive and parallel BH approach to identify both the global and competing local minima within a few hundred iterations. Beyond serving as validation for the algorithm, such configurations provide reliable starting points for further analyses, including symmetry classification, motif identification, and refinement with higher-level electronic-structure methods such as density functional theory (DFT).

In this brief report, we provide a concise overview of the Basin Hopping method, emphasizing its foundational principles. We also discuss strategies to further accelerate the algorithm when applied to first-principles calculations. We believe that these insights will be valuable for students and researchers in the computational chemistry and physics communities.

\section{Conclusions}

We have presented an adaptive and parallel implementation of the Basin Hopping algorithm tailored for Lennard-Jones clusters. By combining dynamic control of perturbation magnitude with simultaneous evaluation of multiple trial structures, the method efficiently explores complex potential energy surfaces and reliably identifies low-energy configurations. Our results demonstrate that near-linear acceleration can be achieved when up to eight candidates are evaluated concurrently per BH step, providing a practical strategy to reduce wall-clock time while maintaining the quality of the search. Beyond this point, additional candidates offer diminishing returns due to synchronization and communication overhead. This implementation establishes a robust foundation for future extensions, including multithreaded systems, machine-learned potentials (e.g., neural networks trained to predict cluster energies from atomic coordinates, such as graph neural networks), empirical force fields, integration with first-principles methods, and hybrid optimization strategies leveraging surrogate models.


\section{Acknowledgments}
O.C. gratefully acknowledges the Secretaría de Ciencia, Humanidades, Tecnología e Innovación (SECIHTI) of Mexico for the master’s grant provided under CVU 2056827. P.L.R.-K. would like to thank the support of CIMAT Supercomputing Laboratories of Guanajuato and Puerto Interior. We acknowledge the use of Python open-source scientific libraries and the computational support from local clusters.


\appendix
\section*{Appendix: Source Code (\texttt{basin\_adapt.py})}

The complete Python source code implementing the adaptive and parallel Basin Hopping algorithm is included directly in the \LaTeX\ source of this submission. The main script, \texttt{basin\_adapt.py}, contains the implementation used to generate all results discussed in this work.





\bibliographystyle{unsrt}
\bibliography{mendelei.bib}
\end{document}